\documentclass[fleqn,usenatbib]{mnras}

\usepackage{newtxtext,newtxmath}
\usepackage[T1]{fontenc}

\DeclareRobustCommand{\VAN}[3]{#2}
\let\VANthebibliography\thebibliography
\def\thebibliography{\DeclareRobustCommand{\VAN}[3]{##3}\VANthebibliography}

\usepackage{graphicx}	% Including figure files
\usepackage{amsmath}	% Advanced maths commands
\usepackage{amssymb}	% Extra maths symbols

\title[Giant pulses from J1823$-$3021A]{Giant pulses from J1823$-$3021A observed with the MeerKAT telescope}

\author[F. Abbate et al.]{
F.~Abbate,$^{1,2}$\thanks{E-mail: abbate@mpifr-bonn.mpg.de},
M.~Bailes$^{3,4}$,
S.~J.~Buchner$^{5}$
F.~Camilo$^{5}$,
P.~C.~C.~Freire${^1}$,
M.~Geyer$^{5}$, \newauthor
A.~Jameson$^{3,4}$
M.~Kramer$^{1}$, 
A.~Possenti$^{2,6}$,
A.~Ridolfi$^{2,1}$,
M.~Serylak$^{5,7}$,
R.~Spiewak$^{3,4}$, \newauthor
B.~W.~Stappers$^{8}$, 
V.~Venkatraman~Krishnan$^{1}$
\\
$^{1}$ Max Planck Institut f\:ur Radioastronomie, Auf dem Hugel 69, Bonn, Germany\\
$^{2}$ INAF, Osservatorio Astronomico di Cagliari, Via della Scienza 5, I-09047 Selargius (CA), Italy \\
$^{3}$ Centre for Astrophysics and Supercomputing, Swinburne University of Technology, PO Box 218, Hawthorn, VIC 3122, Australia\\
$^{4}$ ARC Centre of Excellence for Gravitational Wave Discovery (OzGrav), Mail H29, Swinburne University of Technology, PO Box 218, Hawthorn, \\ VIC 3122, Australia \\ 
$^{5}$ South African Radio Astronomy Observatory, 2 Fir Street, Black River Park, Observatory 7925, South Africa\\
$^{6}$ Universit\'a di Cagliari, Dipartimento di Fisica,  S.P. Monserrato-Sestu Km 0,700,  I-09042 Monserrato (CA), Italy \\
$^{7}$ Department of Physics and Astronomy, University of the Western Cape, Bellville, Cape Town 7535, South Africa \\
$^{8}$Jodrell Bank Centre for Astrophysics, Department of Physics \& Astronomy, The University of Manchester, Manchester M13 9PL, UK
}

\date{Accepted XXX. Received YYY; in original form ZZZ}

\pubyear{2020}

\begin{document}
\label{firstpage}
\pagerange{\pageref{firstpage}--\pageref{lastpage}}
\maketitle

% Abstract of the paper
\begin{abstract}
The millisecond pulsar J1823$-$3021A is a very active giant pulse emitter in the globular cluster NGC 6624. New observations with the MeerKAT radio telescope have revealed 14350 giant pulses over 5 hours of integration time, with an average wait time of about 1 second between giant pulses. The giant pulses occur in phases compatible with the ordinary radio emission, follow a power-law distribution with an index of $-2.63 \pm 0.02$ and contribute 4 percent of the total integrated flux. The spectral index of the giant pulses follows a Gaussian distribution centered around $-1.9$ with a standard deviation of 0.6 and is on average flatter than the integrated emission, which has a spectral index of $-2.81\pm 0.02$. The waiting times between the GPs are accurately described by a Poissonian distribution, suggesting that the time of occurrence of a GP is independent from the times of occurrence of other GPs. 76 GPs show multiple peaks within the same rotation, a rate that is also compatible with the mutual independence of the GP times of occurrence. We studied the polarization properties of the giant pulses finding, on average, linear polarization only at the 1 percent level and circular polarization at the 3 percent level, similar to the polarization percentages of the total integrated emission. In 4 cases it was possible to measure the RM of the GPs which are highly variable and, in two cases, is inconsistent with the mean RM of the total integrated pulsar signal.
\end{abstract}

% Select between one and six entries from the list of approved keywords.
% Don't make up new ones.

%%%%%%%%%%%%%%%%%%%%%%%%%%%%%%%%%%%%%%%%%%%%%%%%%%

%%%%%%%%%%%%%%%%% BODY OF PAPER %%%%%%%%%%%%%%%%%%

\section{Introduction}

Giant pulses (GPs) are individual pulses from a pulsar that are significantly stronger than the average pulse. They were initially discovered in the Crab pulsar \citep{Staelin1968} but they have been observed in different pulsars \citep{Knight2006review} including some millisecond pulsars (MSPs) like B1937+21 \citep{Cognard1996}, J0218+4232 \citep{Knight2006b}, B1957+20 \citep{Knight2006b}, and J1824$-$2452A \citep{Romani2001, Knight2006}.

The energetics of ordinary pulsar emission usually follow either a Gaussian or a log-normal distribution (e.g. \citealt{Burke-Spolaor2012}) and it is rare to observe single pulses more than a few times stronger than the average pulse. In contrast, GPs are described by a power-law distribution which declines more slowly than a log-normal and therefore high-energy pulses with fluences more than ten times the mean are much more common \citep{Lundgren1995,Johnston2002}. In this work we will refer to a pulse as a GP if it lies on the power-law distribution.
Apart from the power-law distribution, GPs are characterised by small intrinsic widths of a few microseconds or less and their occurrence only in specific rotational phases.
In MSPs, GPs usually do not occur at the same phases as the ordinary radio emission, but at the trailing edge of the radio profile components and are typically coincident with X-ray emission \citep{Romani2001,Knight2006b}. Because of this, it was suggested that GPs originate in the same region as the high energy emission and that GPs could be the radio component of that emission \citep{Cusumano2003}.
However, the sample of GP-emitting MSPs is still too small to draw definitive conclusions. GPs also have different spectral properties than the ordinary radio emission. In PSR B1937+21, GPs have steeper spectrum than the ordinary emission \citep{Kinkhabwala2000}.

Polarimetric studies of GPs in MSPs show that in PSR B1937+21, J1824$-$2452A, and J1957+20 the GPs are significantly polarized, differently from the ordinary emission, with some GPs reaching 100 percent polarisation \citep{Knight2006, McKee2019, Li2019}.

The arrival times of GPs in the Crab pulsar show a Poissonian distribution \citep{Lundgren1995, Karuppusamy2010} suggesting that the emission is neither suppressed nor enhanced by the presence of other GP activity. The GPs of PSR B1937+21 follow the same arrival time statistics \citep{McKee2019}, while in PSR J1957+20 the GPs are better described by a Weibull distribution and might be linked to the changing of the mode of emission \citep{Mahajan2018}. 

%theoretical explanations
There are different theories explaining the mechanism behind the generation of GPs in pulsars. Some models predict that the emission arises near the light cylinder of the pulsar and is caused by perturbations in the electron-positron plasma \citep{Istomin2004,Lyutikov2007, Machabeli2019}. In these models, GPs are discharge events in which the energy accumulated in the plasma is released in the form of radio waves.
A different model predicts that GPs are caused by induced Compton scattering of the ordinary radio emission in the ultra-relativistic and highly magnetized plasma of the magnetosphere \citep{Petrova2004b, Petrova2004}. These scattering events can focus the emission onto much smaller opening angles while boosting the flux. Other theoretical models are described in \cite{Cairns2004}.

The pulsar J1823$-$3021A is found near the center of the Galactic globular cluster NGC6624 at a distance of $\sim 7.4$ kpc from the Sun \citep{Baumgardt2019}\footnote{The updated properties of the globular cluster can be found at the webpage \url{https://people.smp.uq.edu.au/HolgerBaumgardt/globular/orbits.html}}. This pulsar has a period of 5.44 ms \citep{Biggs1994} and is visible as a gamma-ray pulsar with high energy pulses coincident with the radio pulses \citep{Freire2011}. It has one of the highest inferred spin-down rates of all MSPs. It has been suggested that the observed period derivative is heavily influenced by the gravitational potential of the cluster NGC 6624 and that it could be a sign of the presence of an intermediate mass black hole in the centre of the cluster \citep{Perera2017a}. However, the strong rotation-powered gamma-ray flux suggests that a significant portion of the spin down is intrinsic to the pulsar \citep{Freire2011}. The very high spin-down rate is not unique among MSPs: PSR J1824$-$2452A, another GP emitter, also has a very large spin-down rate and strong gamma-ray emission \citep{Johnson2013}.

This pulsar is known to emit GPs \citep{Knight2005, Knight2007} occurring roughly at the same rotational phases as the radio and gamma-ray pulsations. Previous studies with the Parkes radio telescopes revealed 120 GPs in two coherently de-dispersed observations each 9000 s long in a frequency band 64 MHz wide centred around 685 MHz. The time resolution in these observations was 16 $\mu$s \citep{Knight2007}.
The energies of the observed GPs follow a power-law distribution with a slope of $-3.1$ and the arrival times of the GPs appear to follow a Poisson distribution. 

We report the results of observations of PSR J1823$-$3021A in L-band (963–1605 MHz) with the MeerKAT radio telescope in South Africa as part of the MeerTIME\footnote{\url{http://www.meertime.org}} large science project \citep{Bailes2018}.
This telescope has a very low system equivalent flux density (SEFD) ($\sim7$Jy) when compared to the SEFD  of 66 Jy for the reported Parkes observations \citep{Knight2007}, a large frequency band (642 MHz) and a higher time resolution of 10$\mu$s. Therefore, we expect to observe a much larger number of GPs from this pulsar and study its properties in greater details.

\section{Observations and Analysis}

%Meerkat
PSR J1823$-$3021A was observed with the MeerKAT radio telescope in South Africa \citep{Jonas2009,Booth2012} with the pulsar timing user supplied equipment (PTUSE) backend \citep{Bailes2020}. The backend recorded search mode (psrfits format, \citealt{Hotan2004}) data across 642 MHz of bandwidth. The observations were performed at L-band centred around 1284 MHz and all Stokes parameters were recorded. More details about the data acquisition and system description can be found in \cite{Bailes2020}.The pulsar was observed for 150 minutes each on the 10th of September 2019 and on the 9th of November 2019. 
In order to create a tied-array beam of the telescope large enough to observe also the other pulsars in the same cluster, only the antennas in the inner kilometer of the array were selected for the observations. In the first observation 36 antennas were used while in the second observation the number of antennas used was 42. The lower number of antennas used reduces the total collecting area of the telescope and, in turn, increases the SEFD quoted in the introduction by a factor of 1.77 for the first observation and by a factor of 1.52 for the second.
The observations were coherently de-dispersed at the known dispersion measure of J1823$-$3021A (86.890 pc cm$^{-3}$) and  were calibrated for polarization offline. This was done by utilising data collected during the array calibration phase. In short, following successful phase up there remains a phase difference between polarizations which is corrected for by using observations of the noise diode and a known polarisation calibrator. This difference is corrected for by applying the solutions to the data using \texttt{pac -Q} in \textsc{psrchive}. More details on pulsar polarimetry using MeerKAT can be found in \citet{Bailes2020} and \citet{Serylak2020}.

%GP detection algorithm.
The data were divided in single-pulse archives using \textsc{dspsr} \citep{vanStraten2011} and a recent ephemeris for PSR J1823$-$3021A. The profiles were created with 512 phase bins across the pulse period with a time resolution of 10 $\mu$s for each bin. The single pulses were subsequently cleaned from radio frequency interference (RFI) using an algorithm that explores the statistics of the off-pulse region on a channel-by-channel basis to search for significant deviations from a normal distribution and deletes the channel when more than 3.5 sigma from the median\footnote{\url{https://github.com/mshamohammadi/rfihunter}}. 
After this process, a copy of the single pulse archives was averaged in frequency to compute the signal-to-noise ratio (S/N) of the total intensity (Stokes I parameter). This computation was performed using the \texttt{pdmp} option within \texttt{psrstat -c snr} of \textsc{psrchive} \citep{Hotan2004,vanStraten2012}. 
The single pulses which showed a S/N higher than 7 were selected. We compute the energy of the average radio pulse by dividing the total S/N of the integrated profile by the square root of the number of rotations observed and we obtain an average S/N of 0.5. Since it is extremely unlikely that single pulses with S/N higher than 7 originate from ordinary emission described by a log-normal distribution, we consider all of the detected pulses as GPs.
A subset of the selected pulses was visually analysed to verify that all narrow-band RFI was removed.

\section{Results}

\begin{figure}
\centering
  \centering
  \includegraphics[width=0.9\linewidth]{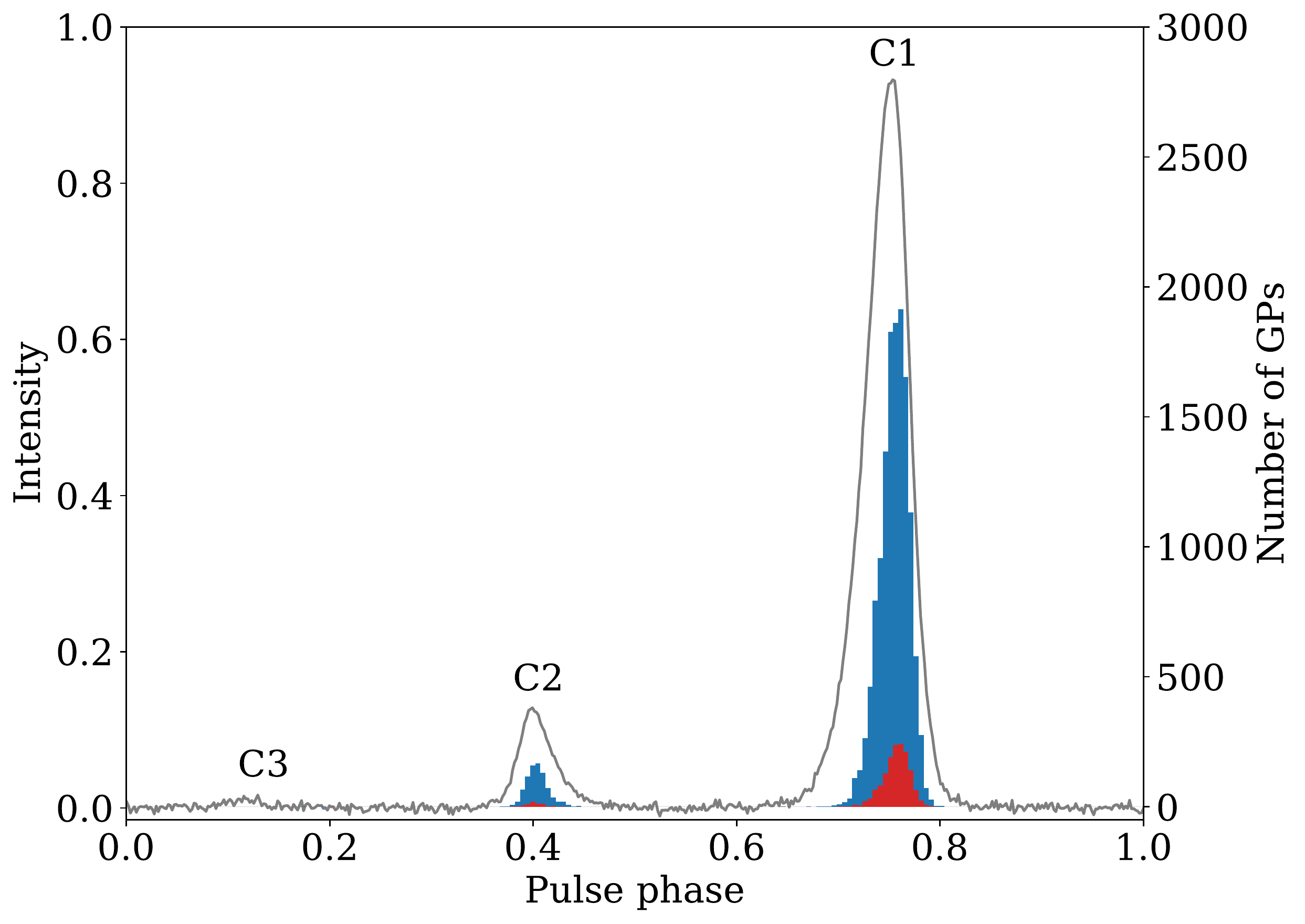}
  \caption{Histogram showing the phase distribution of the detected GPs. The red histogram shows the phase distribution of the brightest 10 percent GPs. The integrated radio profile of PSR J1823$-$3021A is shown in grey. The data are shown as relative intensity and normalised so that the peak of the average pulse profile has a value of 1.0. }
  \label{fig:phase_distribution}
\end{figure}

\begin{figure}
\centering
  \centering
  \includegraphics[width=0.9\linewidth]{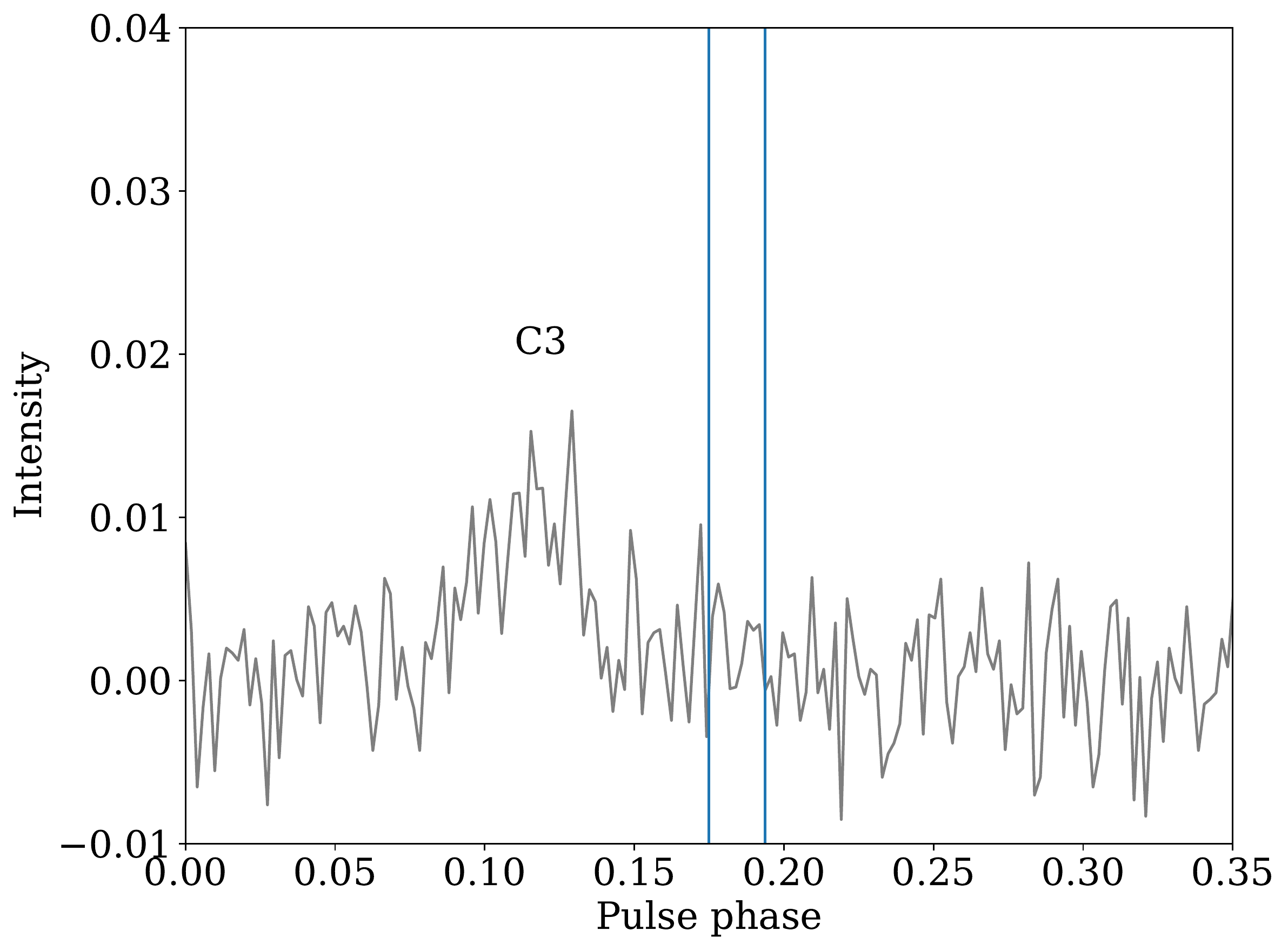}
  \caption{A zoomed in view of the weak third component showing the location where two GPs were detected.}
  \label{fig:third_pulse_phase}
\end{figure}

A total of 14350 GPs with S/N higher than 7 were detected in the 18000s of observation averaging about 1 GP every 1.25s or 1 GP every 230 rotations of the pulsar. As previously noted in \citep{Knight2007}, the GPs fall within the pulse phases of the regular pulsar emission. Figure \ref{fig:phase_distribution} shows the histogram of the phases of the detected GPs superimposed on the integrated radio emission of the pulsar. A total of 13522 GPs fall within the main component of the integrated emission C1 with an average of 1 every 1.3s, while only 828 fall within the component C2 with an average of 1 every 21s. The GPs in the main component C1 mostly occur in the trailing edge of the component. The window of occurrence has a small dependence on the energy of the GPs as shown in the red histogram of Figure \ref{fig:phase_distribution}. This histogram shows the phase distribution of only the brightest 10 percent of the GPs and is skewed towards the trailing side of component C1. A Kolmogorov-Smirnov test shows that the phase distribution of the brightest GPs is different from that of the weakest GPs (p-value of the null-hypothesis < $10^{-3}$).
As shown in Figure \ref{fig:third_pulse_phase}, two more GPs were detected close in phase to a previously undetected component of the integrated pulse profile, C3. Visual inspection of these two GPs confirms that they show emission across the entire band at the dispersion measure of the pulsar.
These two GPs have a delay of $\sim 0.06$ pulse phases with respect to the component C3. Even though the numbers are too small to derive the full range of phases that emit GPs, similar delays have been observed in other MSPs like PSR J1824$-$2452A \citep{Knight2006, Bilous2015} and J0218+4232 \citep{Knight2006b}.

\begin{figure}
\centering
  \centering
  \includegraphics[width=0.9\linewidth]{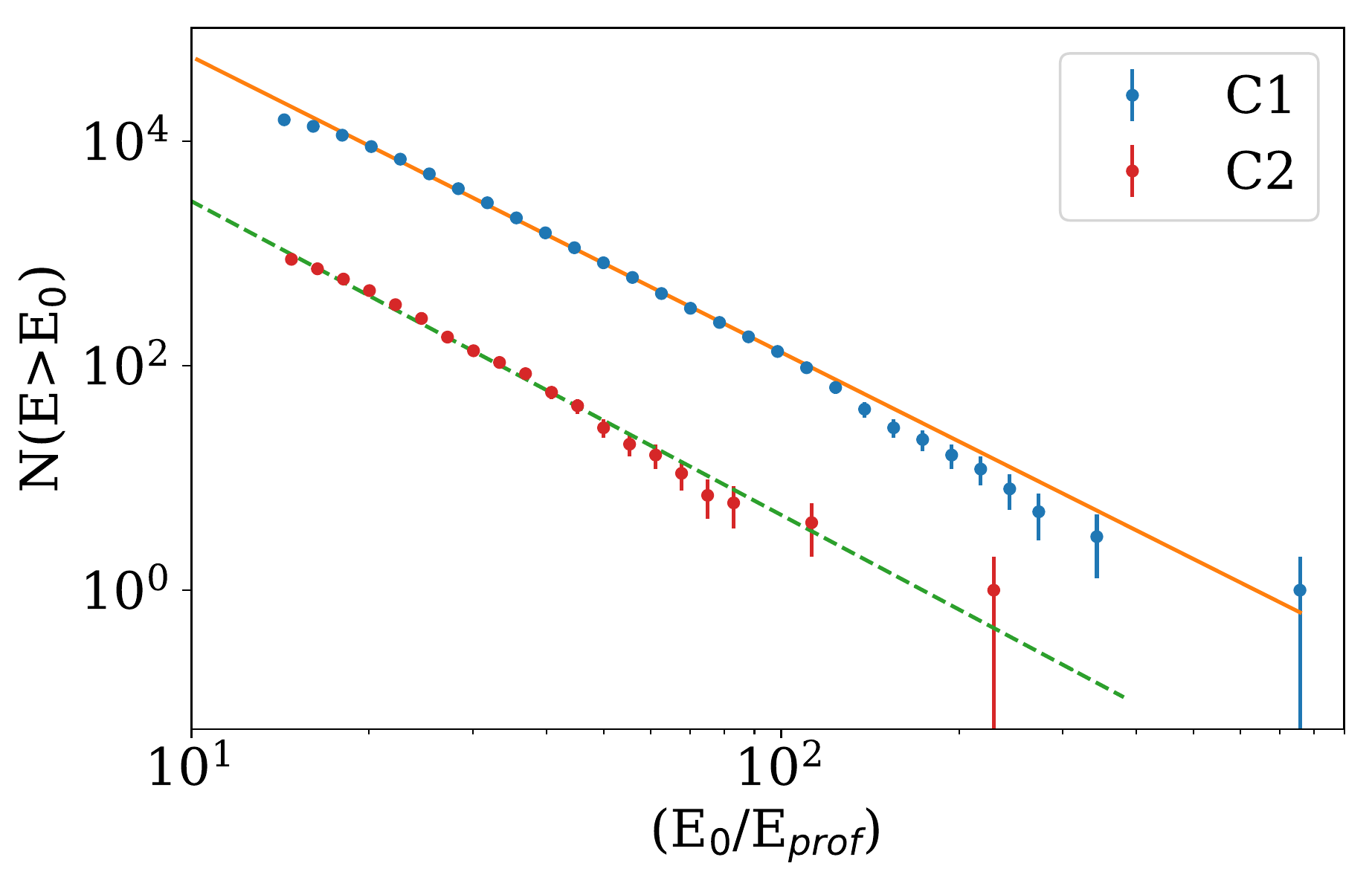}
  \caption{Cumulative distribution of the GP energies in the different components normalised by the average pulse energy. The best fit power law to the GPs in C1 (orange line) has an index of $-2.63 \pm 0.02$ while for the GPs in C2 the best fit power law (green dashed line) has an index of $-2.79 \pm 0.04$. }
  \label{fig:pulse_energy}
\end{figure}

%pulse energies
The energies of the GPs are consistent with a power-law distribution as shown in Figure \ref{fig:pulse_energy}. The GPs in the two components C1 and C2 are treated separately. The brightest pulse detected has a S/N of $\sim$ 370.
The x-axis in Figure \ref{fig:pulse_energy} shows the ratio of the flux density of the detected GPs divided by the flux density of the average radio pulse.  The best power-law index that describes the distribution is found using a Bayesian chi-square minimization algorithm based on the \textsc{EMCEE} python package \citep{Foreman-Mackey2013} and, for the GPs in C1 is $-2.63\pm 0.02$, while for the GPs in C2 is $-2.79 \pm 0.04$. Repeating the fits excluding the 10 brightest GPs, we conclude that the power-law indices are not heavily influenced by the high-energy tail of the distributions.
This power-law index is less steep than the value estimated by \cite{Knight2007} of $-3.1$. This difference could be related to the slight steepening we observe at higher energies.
This result confirms that all of the pulses selected as GPs lie on a power-law distribution.

We estimate the contribution of the GPs to the integrated pulse emission by adding all the detected GPs together and find that only 4 percent of the emission in C1 is due to GPs and 2 percent of the emission in C2 is due to GPs.
These percentages are measured using the GPs that have positively passed the detection algorithm meaning that they have a S/N > 7. However, there might be a large population of undetected GPs below this limit.
We explore the possibility that the GP energy distribution continues to lower energies with the same power-law index. If we sum all of the GPs with energy twice as large as the average energy per pulse, the flux in GPs would correspond to 100 percent of the total integrated flux. As a consequence, if the emission were entirely made up of pulses following a power-law distribution, there must be a cut off in the energy distribution around this level. The shape of the component C1 obtained using only GPs does not account completely for the shape of the integrated profile. However, since the window of occurrence of the GPs appears to be slightly dependent on the pulse energy, the pulse shape of weaker GPs is not known. Nevertheless, the ratio between the GP flux in component C1 and C2 is twice as that of the integrated profile. Therefore, the integrated profile is not likely to be entirely made up of GPs and the energy distribution observed in Figure \ref{fig:pulse_energy} should have a break in the power-law at energies lower than the observed GPs to transition to the ordinary log-normal emission.

\begin{figure}
\centering
  \centering
  \includegraphics[width=0.9\linewidth]{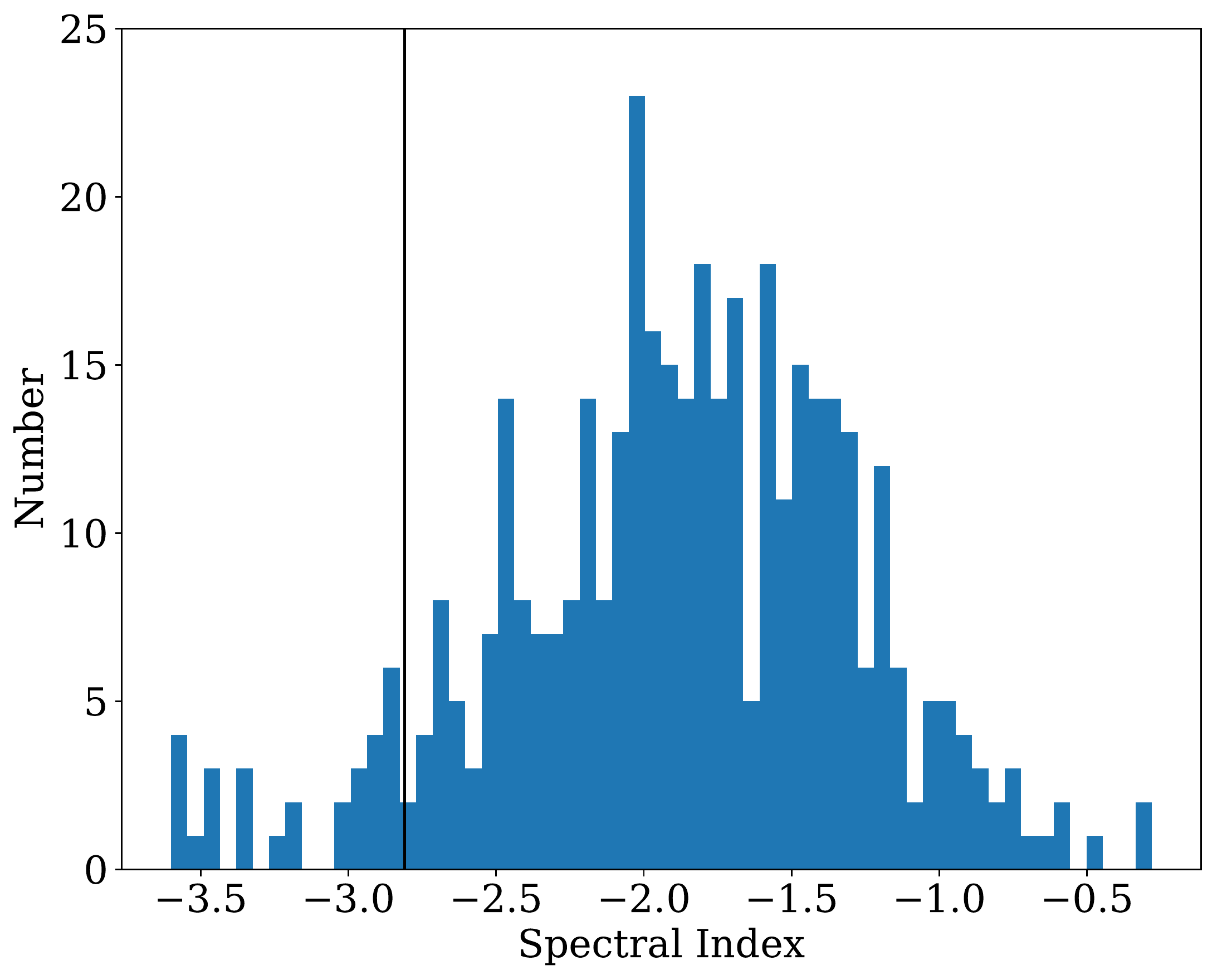}
  \caption{Histogram showing the spectral indices of the 400 brightest GPs. The black line shows the spectral index of the integrated profile. }
  \label{fig:spectral_index}
\end{figure}

%spectral index
The emission of the GPs extends over the entire observing bandwidth and can be modelled with a single power-law. An example of typical spectrum is shown in Figure \ref{fig:stokes_spectrum}.
This frequency behaviour allows us to measure the spectral index of the brightest 400 GPs. The resulting histogram is shown in Figure \ref{fig:spectral_index}. The measured spectral indices appear to follow a Gaussian distribution centered around $-1.9$ with a standard deviation of 0.6. The average error on the spectral indices of the GPs is 0.09, significantly smaller than the standard deviation of the distribution. The spectral index of the integrated emission (black line in the Figure) is $-2.81\pm 0.02$. This discrepancy is significant given that an important percentage of the integrated emission could be in the form of GPs. The spectral index of the normal emission, not considering the GPs, would be even lower than what is measured.

\subsection{Arrival time statistics}

\begin{figure}
\centering
  \centering
  \includegraphics[width=0.9\linewidth]{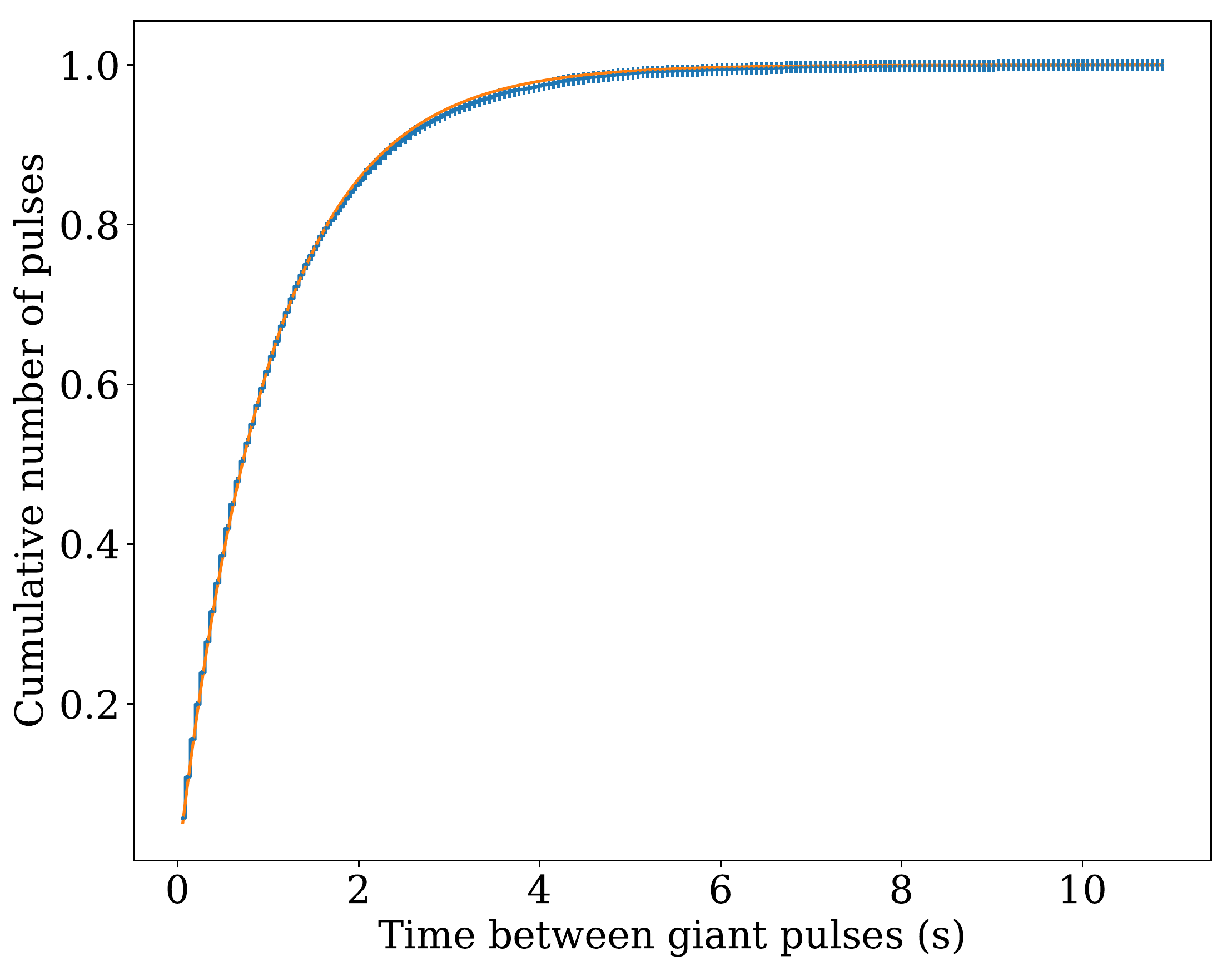}
  \caption{Normalized cumulative distribution of the time interval between GPs. Each bin in the histogram spans 54\,ms or 10 rotations. The distribution is fitted with an exponential with decay time of 1s. This shows that the time of arrival of GPs is well described by a Poisson distribution.}
  \label{fig:wait_times}
\end{figure}

Studying the arrival times of the GPs provides information on the origin of this phenomenon. \cite{Knight2007} suggested that in PSR J1823$-$3021A the emission times of the GPs are well modelled by a Poissonian distribution. In this distribution, the time interval between GPs follows a purely exponential decay and this implies that the GPs occur independently from one another. The larger number of observed GPs allows us to study the statistic of the arrival times in greater detail. We show the cumulative distribution of time intervals between GPs measured in our observations in Figure \ref{fig:wait_times} together with the best-fitting exponential distribution.
The histogram is binned into 10 rotations. The errors on the number of pulses in every bin are estimated as the square root of the number and is compatible with the exponential fit. A Kolmogorov-Smirnov test shows that the observed distribution is compatible with the best-fitting model (p-value of the null hypothesis $>0.05$). The 68 percent confidence interval of the time decay of the best-fitting exponential is $(1.0086, 1.0092)$ s. This suggests that the arrival times of the GPs follow a Poissonian distribution.

\begin{figure}
\centering
  \centering
  \includegraphics[width=0.9\linewidth]{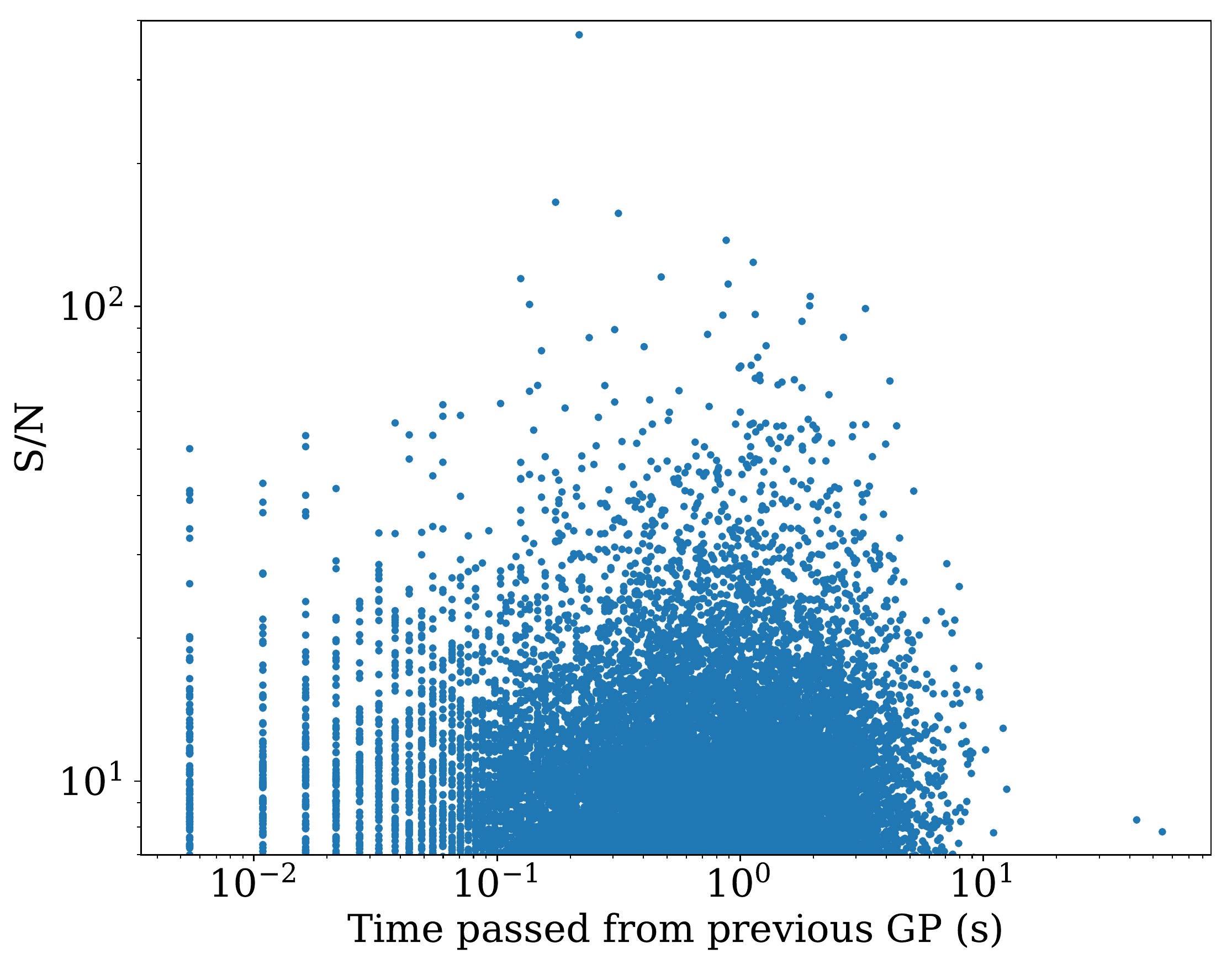}
  \caption{Value of the S/N of each GP plotted against the time that has elapsed from the previous GP. The time interval between GPs is always a multiple of the rotation of the pulsar
  which is about 5.4 ms.}
  \label{fig:wait_times_snr}
\end{figure}

The value of the S/N of each GP plotted against the amount of time that elapsed from the previous GP is shown in Figure \ref{fig:wait_times_snr}. There is no evident correlation between the two quantities.
This suggests that the intensity of a GP does not depend on the total amount of energy that can be accumulated in the magnetosphere of the pulsar in the time that passed from the previous GP.  

\subsection{Giant pulses with multiple peaks}

\begin{figure}
\centering
  \centering
  \includegraphics[width=0.9\linewidth]{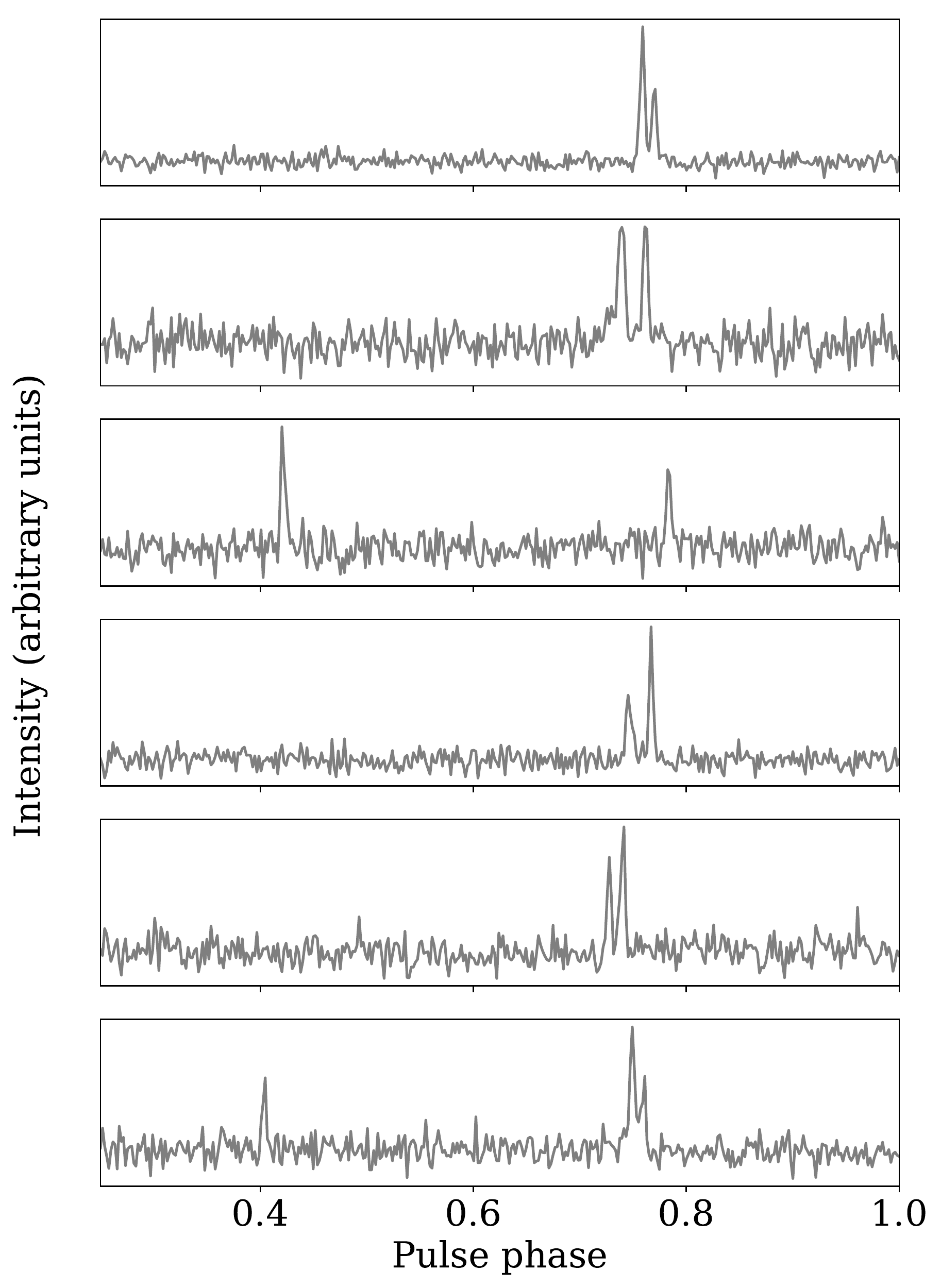}
  \caption{6 Examples of GPs showing multiple components in a single rotation.}
  \label{fig:multiple_components_example}
\end{figure}

A significant number of GPs have profiles that are composed of multiple peaks. A total of 76 GPs show a second peak with an S/N higher than 7. We make sure that both peaks pass the S/N limit by repeating the measure of S/N after having subtracted the other peak.
A few examples are shown in Figure \ref{fig:multiple_components_example}. The pulse phases of these plots matches that in Figure \ref{fig:phase_distribution}.  
We classify as double-peaked GPs only the ones where the peaks are separated by more than two phase bins corresponding to 20 $\mu$s. 
In 74 cases the peaks appear in the main component C1 of the integrated pulse profile and in 2 cases the peaks are aligned with the different components C2 and C1. A single GP with emission in both components of the pulse profile had been previously observed by \cite{Knight2007}. In the last panel of Figure \ref{fig:multiple_components_example} we show the only detected case where three distinct peaks are visible, two in the component C1 and one in the component C2.

In this analysis, we counted the rotations of the pulsar starting from the pulse phase 0 shown in Fig. \ref{fig:phase_distribution}. Because of this, the GPs with emission in different components that we detected are only those where the peak in component C2 precedes the one in component C1. If we had chosen to start counting the rotations of the pulsar starting from the pulse phase 0.5 of Fig. \ref{fig:phase_distribution}, we would have considered double-peaked GPs the ones where the emission in component C1 precedes the emission in component C2. The number of GPs that show this behaviour is 4.

If we assume that the multiple peaks are different GPs independent from one another, we can estimate the probability of occurrence of two GPs in the same rotation. The probability of occurrence of a single GP in any rotation in the main component of the pulse profile (C1 in Figure \ref{fig:phase_distribution}), called P(C1), is $1/244$. The probability of having two GPs in the same rotation is $\textrm{P(C1)P(C1)}=1/59536$. Given that the total number of rotations present in the observations is $3308823$, we should expect 56 different cases with two GPs in the component C1 compatible at the 2$\sigma$ level with the observed number of 74.

The probability of occurrence of a single GP in any rotation in the secondary component of the pulse profile (C2 in Figure \ref{fig:phase_distribution}), called P(C2), is $1/3996$. The probability of having one GP in C1 and one in C2 in the same rotation is $\textrm{P(C1)P(C2)}=1/975024$. We should expect 3 different rotations with GPs both in the component C1 and C2. This is compatible both with the 2 detected in which the component C2 precedes the component C1 and the 4 in which the component C2 follows the component C1.

The probability of observing one case in which three peaks are visible in a single rotation, calculated with the prescription above, is only 1.4 percent. This probability is very low, but still consistent with the observation at 2.5$\sigma$.

\begin{figure}
\centering
  \centering
  \includegraphics[width=0.9\linewidth]{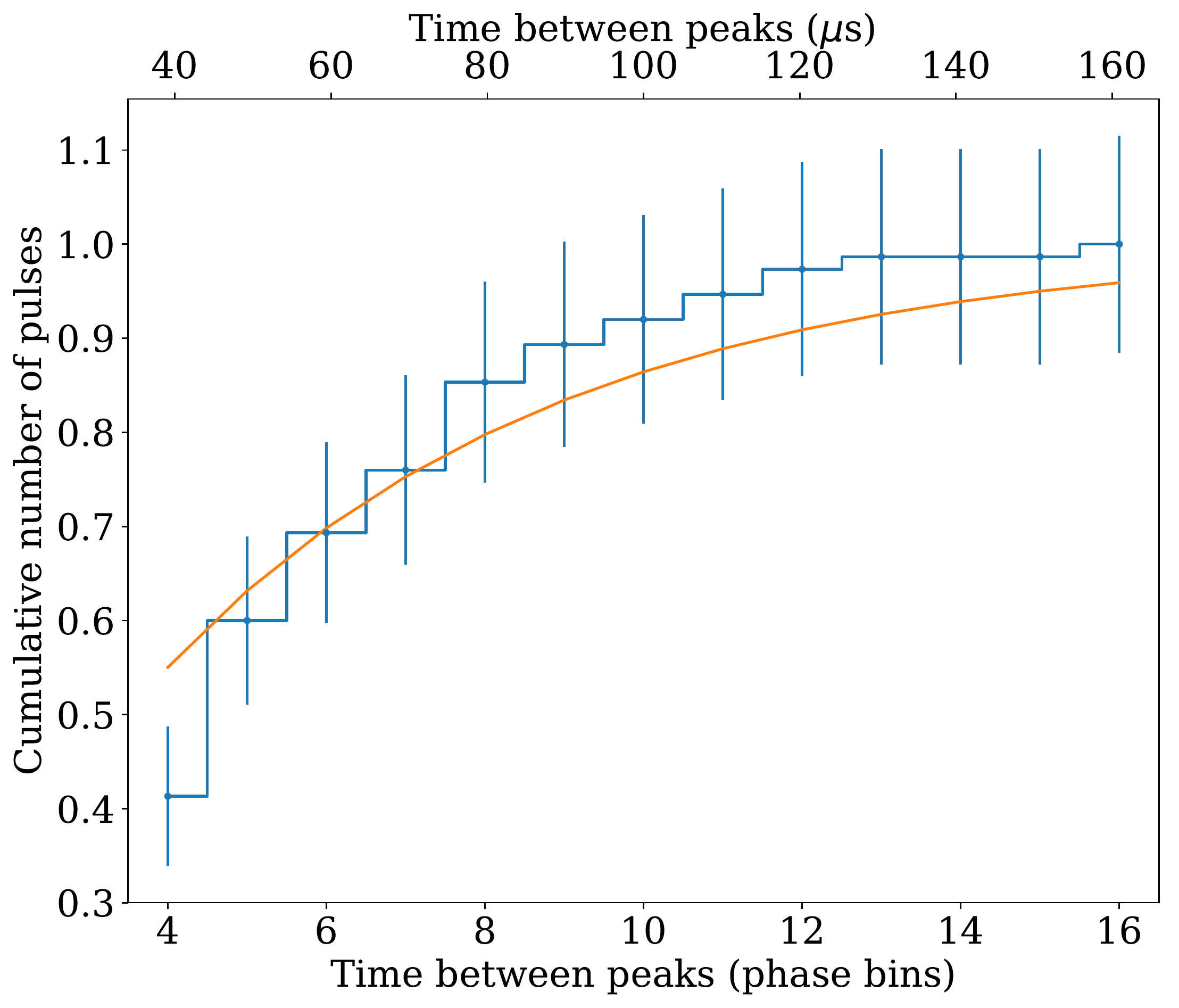}
  \caption{Normalized cumulative distribution of the phase difference between GPs occurring during the same rotation. The phase difference is reported in both number of phase bins (lower) and in $\mu$s (upper). Each phase bin has a time resolution of 10 $\mu$s. The distribution is fitted with an exponential.}
  \label{fig:multiple_components_analysis}
\end{figure}

If we consider the multiple peaks as different GPs occurring at different rotational phases in the same rotation, we can analyse the difference in phase between the peaks. Repeating the analysis as before, we check if they can be described by a Poissonian distribution. Figure \ref{fig:multiple_components_analysis} shows the cumulative distribution of the phase difference between the different peaks. The best-fitting exponential is also plotted with an orange line. A Kolmogorov-Smirnov test shows that the observed distribution is compatible with the best-fitting model (p-value of the null hypothesis $>0.2$). Converting the exponential decay parameter from phase bins to time, we find a 68 percent confidence interval from 48 $\mu$s to 58 $\mu$s. The compatibility with an exponential distribution (and in turn with an underlying Poisson statistics) suggests that the phases at which the GPs occur are not influenced by the presence of another GP in the same rotation. In some models of the origin of GPs (see the Discussion), a minimum interval of time is required in order to accumulate the energy necessary for emitting a GP. We claim that if this time interval exists, it must be shorter than the time resolution of our observations of 10 $\mu$s.

The double GP shown in the top panel of Figure \ref{fig:multiple_components_example} is bright enough to allow an estimate of the spectral index for both components. The first peak has a spectral index of $-1.6 \pm 0.1$, while the second peak (separated by 6 phase bins) has a spectral index of $-3.0 \pm 0.3$. This shows that large variability in the spectrum of GPs can occur in time scales as short as 60 $\mu$s. 

\subsection{Polarization}

\begin{figure}
\centering
  \centering
  \includegraphics[width=0.9\linewidth]{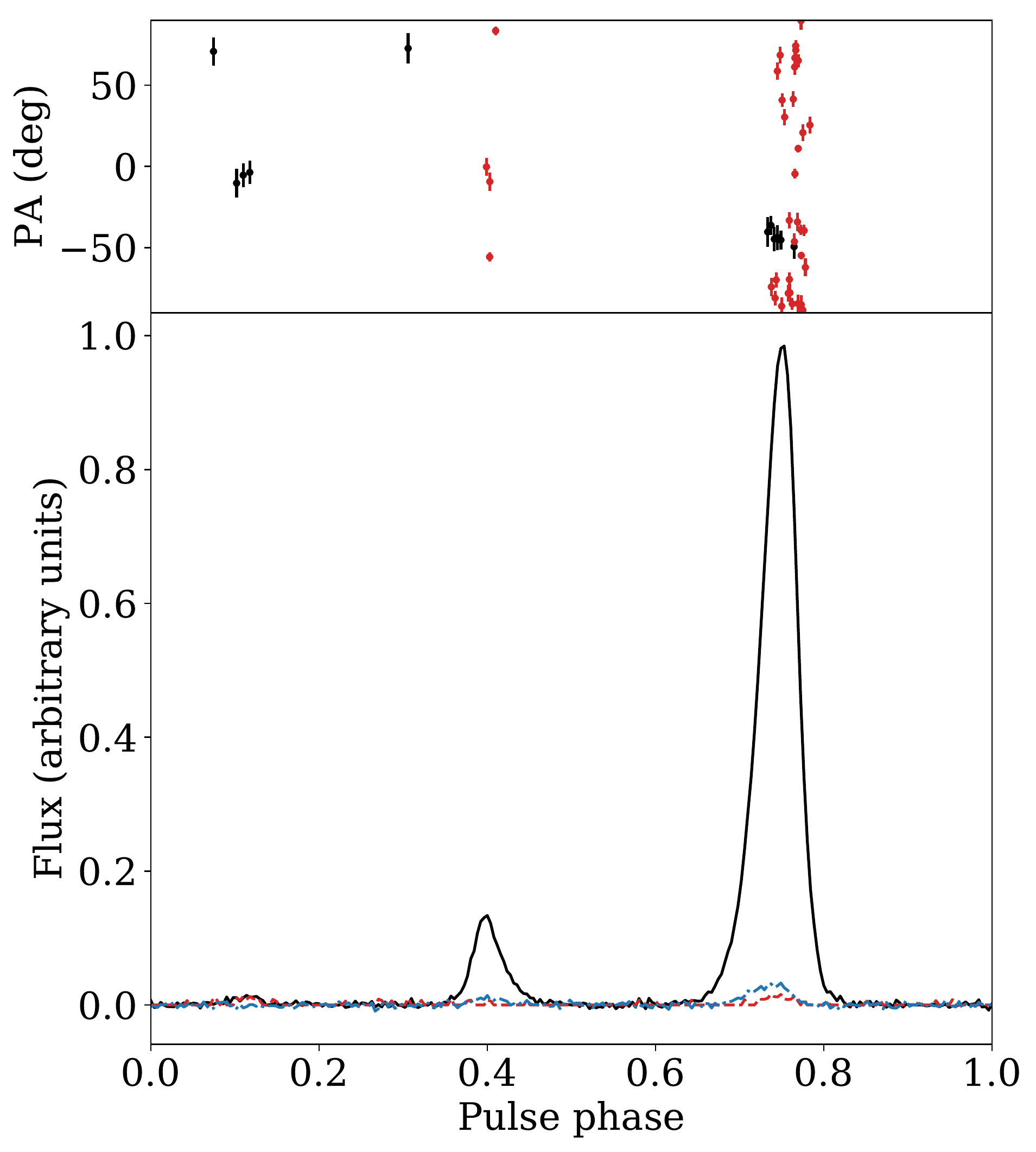}
  \caption{Polarization profile of J1823$-$3021A. In the lower panel, the black line shows the total intensity, the red dashed line shows the linear polarization and the blue dot-dashed line the circular polarization. The panel above shows in black the values of the position angle (PA) measured for the bins where the linear polarization is higher than 3 times the standard deviation of the noise in the off-pulse region. The PAs of the GPs that are significantly linearly polarized are shown in red.}
  \label{fig:polarization_profile}
\end{figure}

The emission of pulsar J1823$-$3021A is not strongly polarized. The integrated polarization profile is shown in Figure \ref{fig:polarization_profile}. The linear polarization has been obtained by summing in quadrature the Stokes parameters U and Q and by removing the positive bias according to the method described in \cite{Wardle1974}.
The components C1 and C2 show linear polarization only at the 1 percent level and circular polarization at the 3 percent level. Despite this, it was possible, thanks to the high S/N of the observation, to estimate the rotation measure (RM) of the pulsar and its $1\sigma$ uncertainty to be $-16 \pm 3$ rad m$^{-2}$.

\begin{figure}
\centering
  \centering
  \includegraphics[width=0.9\linewidth]{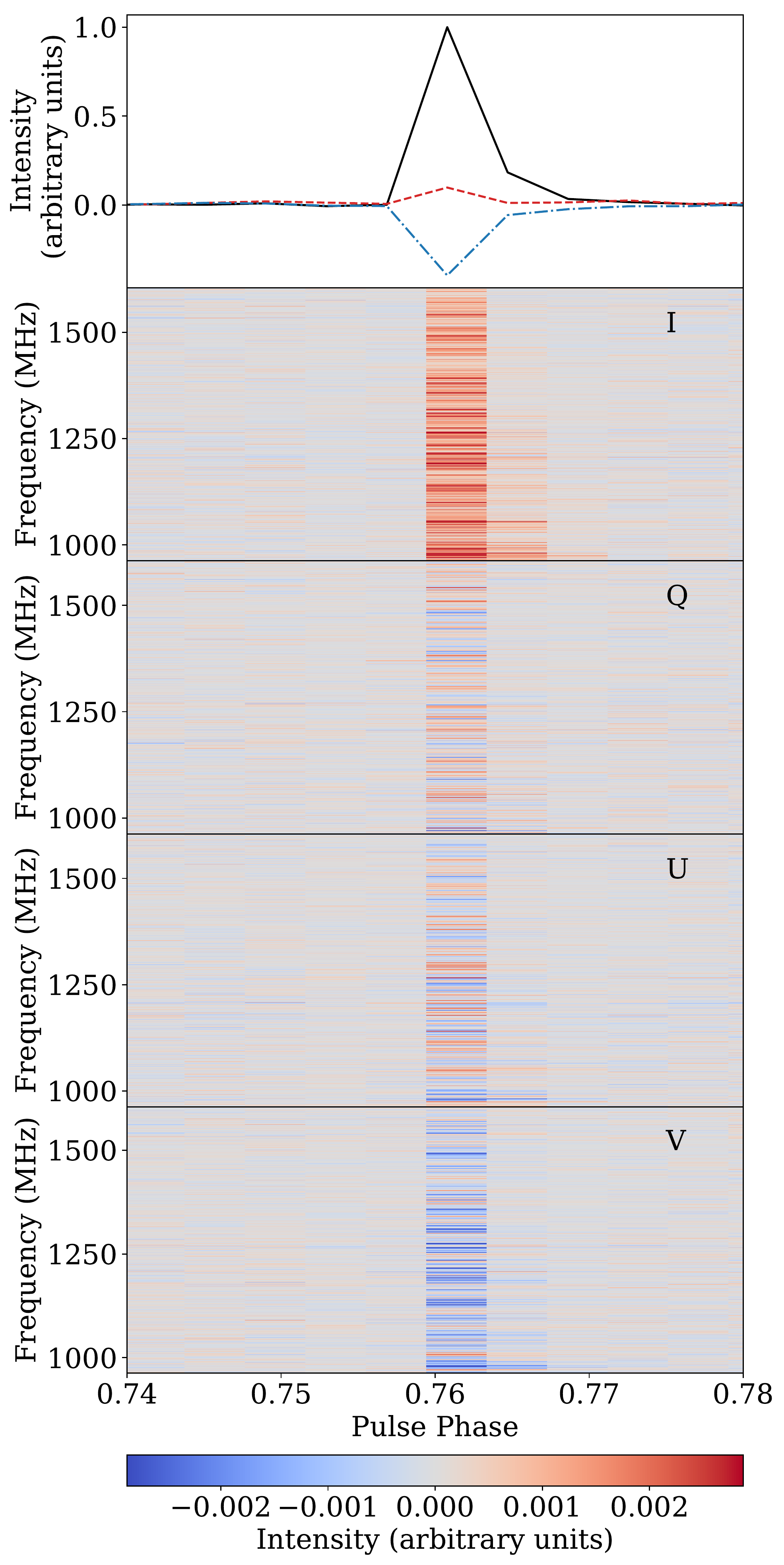}
  \caption{Example of a polarized and bright GP. The top panel shows the polarization profile. The black line shows the total intensity, the red dashed line shows the linear polarization and the blue dot-dashed line the circular polarization. The four bottom panels show the spectrum of the four Stokes parameters across the entire band.}
  \label{fig:stokes_spectrum}
\end{figure}

The polarization properties of the GPs do not vary significantly from those of the total integrated profile in contrast with what has been observed in other MSPs like PSR B1937+21 \citep{McKee2019} and PSR J1824$-$2452A \citep{Bilous2015}. 
The number of GPs that show linear polarized flux stronger than 5 times the noise level is only 39. The position angle (PA) integrated across all frequencies of these GPs is shown in grey in the top panel of Figure \ref{fig:polarization_profile}. The PAs do not show a clear trend with phase.
The number of GPs that show circular polarized flux stronger than 5 times the noise level is 936. The polarization percentages of the profile obtained by summing all the GPs together are comparable to those of the total integrated profile. 
The spectrum of the Stokes parameters for a bright polarized GP is shown in Figure \ref{fig:stokes_spectrum}. The emission extends throughout the entire frequency band at variance for instance with the case of PSR B1821$-$24A \citep{Bilous2015}
where a number of narrow-band patches are seen. The parameters Q and U show rapid sign changes throughout the frequency band.
% RM values of the brightest GPs.
In 4 of the GPs, the polarized signal across the frequency band was high enough to allow a measure of the RM. The measured RM with 1$\sigma$ uncertainty for the brightest GPs are: $-1.6 \pm 5.8$ rad m$^{-2}$, $-27 \pm 6.2$ rad m$^{-2}$, $3.1 \pm 4.5$ rad m$^{-2}$, and $25.2 \pm 2.2$ rad m$^{-2}$. This last value of RM is inconsistent at the 3$\sigma$ level with the total RM of the pulsar suggesting that different effects contribute to the RM of the GPs. 

\begin{figure}
\centering
  \centering
  \includegraphics[width=0.95\linewidth]{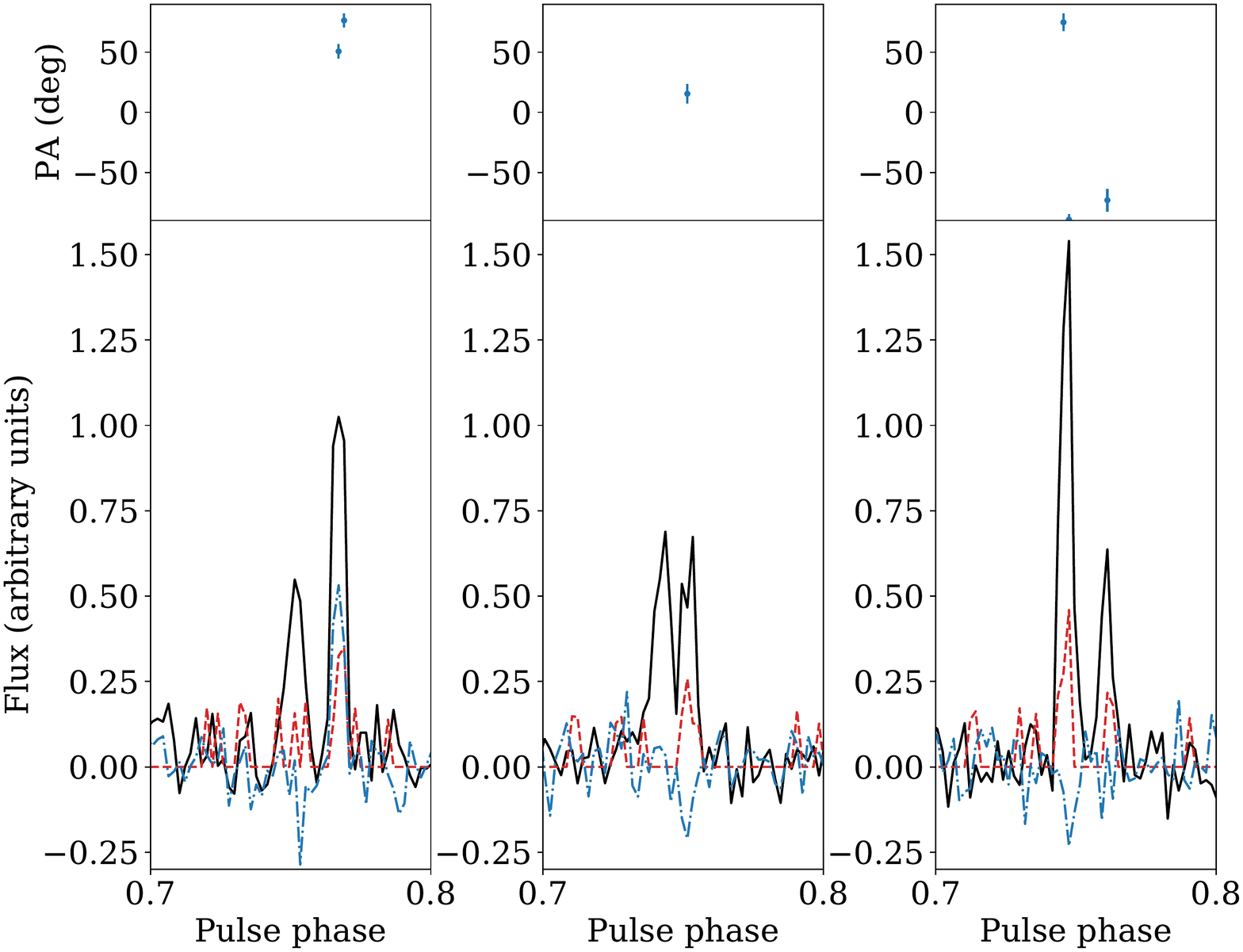}
  \caption{Polarization profiles of three GPs of J1823$-$3021A showing double peaks and different polarization properties in the two peaks. The black line shows the total intensity, the red dashed line shows the linear polarization and the blue dot-dashed line the circular polarization. The panel above shows the values of the position angle (PA) measured for the bins where the linear polarization is higher than 3 times the standard deviation of the noise in the off-pulse region.}
  \label{fig:polarization_profile_GP}
\end{figure}

We analyzed the polarization of multiple GPs that occur within the same rotation. Also for these GPs only a few show significant polarization.  Linear and circular polarizations of the two peaks do not appear to be correlated, indicating that also the polarization properties of the GPs emitted in a same rotation are basically independent.
A collection of double GPs that show significant polarization is shown in Figure \ref{fig:polarization_profile_GP}. The first shows linear polarization only on the second peak and circular polarization in both peaks with opposite signs; the second shows linear and circular polarization only in the second peak while the third shows linear polarization in both peaks and circular polarization only in the first. For the third GP shown in Fig. \ref{fig:polarization_profile_GP}, it is possible to measure the PA in both peaks, and it shows some rotation.

\section{Discussion and conclusions}

We observed the MSP J1823$-$3021A in the globular cluster NGC 6624 with the MeerKAT radio telescope for a total of 5 hours in two different observations and detected 14350 GPs with S/N higher than 7. We observe, on average, 1 GP a second making it the most active known GP emitter amongst MSP. 94 percent of GPs occur at a phase compatible with the main emission component of the pulsar (called C1 in Figure \ref{fig:phase_distribution}). The rest occur simultaneously with the second component C2 with the exception of 2 GPs that are observed slightly delayed from the previously undetected third component of the pulsar C3. The phases of the GPs in C1 and C2 are also coincident with the gamma-ray emission coming from this pulsar. Further gamma-ray observations are necessary to determine whether there is high energy emission coming from the same pulse phases as the GPs close to C3.
The energy distribution of the GPs follows a power-law distribution as previously described by \cite{Knight2007} with a power-law index of $-2.63 \pm 0.02$.

We can compare the number of detected GPs with the previously reported 120 GPs observed with Parkes \citep{Knight2007}.
We first compute the difference in sensitivity between the observations by comparing the SEFD of the telescopes and the frequency bands weighed by the average spectral index of the GPs \citep{Handbook2004}. This increase is:
\begin{equation}
    \frac{A_{\mathrm p}}{A_{\mathrm M}} \frac{N_{\mathrm{ ant}}}{64} \frac{\int_{\nu_{1\mathrm M}}^{\nu_{2\mathrm  M}}\nu^{\alpha} \mathrm{d}\nu }{\int_{\nu_{1\mathrm  P}}^{\nu_{2\mathrm P}}\nu^{\alpha} \mathrm{d}\nu },
\end{equation}
where ${A_{\mathrm p}}$ is the SEFD of Parkes (66 Jy), ${A_{\mathrm M}}$ is SEFD of MeerKAT (7 Jy), $N_{\mathrm{ ant}}$ is the number of MeerKAT antennas used, $\nu_{1 \mathrm M}$ and $\nu_{2\mathrm M}$ are the boundaries of the MeerKAT frequency band, $\nu_{1\mathrm P}$ and $\nu_{2\mathrm P}$ are the boundaries of the Parkes frequency band and $\alpha$ is the average spectral index of the GPs calculated in Figure \ref{fig:spectral_index} to be $-1.9$.
In the first observation the number of antennas used is 36 so this increase is $\sim 17$, in the second observation 42 antennas were used and the increase in sensitivity is $\sim 20$.
This increase in sensitivity leads to an increase of the detected GPs that depends on the energy distribution of the GPs. As estimated in Figure \ref{fig:pulse_energy}, the energy distribution can be modelled with a power law with an index of $-2.63$. The increase in the number of detected GPs is:
\begin{equation}
    \frac{\int_{S_{\mathrm P}}^{\infty} N_0 E^{-2.63} \mathrm{d} E}{\int_{S_{\mathrm M}}^{\infty} N_0 E^{-2.63} \mathrm{d} E} = \left(\frac{S_{\mathrm P}}{S_{\mathrm M}}\right)^{-1.63},
\end{equation}
where $S_{\mathrm P}$ is the sensitivity limit of the Parkes observations, $S_{\mathrm M}$ is the sensitivity of the MeerKAT observations, and $N_0$ is the constant factor of the energy distribution. Averaging between the two MeerKAT observations we expect to observe $126$ times more GPs than in the Parkes observations. This factor is perfectly compatible with the observed increase of $120$. We can conclude that the rate of emitted GPs from J1823$-$3021A did not vary significantly between the observations spaced by 15 years.

%spectral index
The emission of the GPs is broadband covering the entire frequency band of the observation between (963-1603) MHz. This allows us to measure the spectral index of the brightest detected GPs which is, on average flatter, than the total integrated emission. It can be described by a Gaussian distribution centered around $-1.9$ with a standard deviation of 0.6, while the integrated emission has a spectral index of $-2.81 \pm 0.02$. This result is compatible with the model of \cite{Petrova2004} which predicts GPs to have a flatter spectral index with respect to the integrated emission.

Assuming this power-law distribution holds also at energies lower than the cutoff, we estimated that the added flux of GPs would be comparable to the total integrated flux of the pulsar, provided we add together all the extrapolated GPs down to a S/N value higher than 1, in turn corresponding to twice the flux density of the average pulse. Therefore, if the emission is entirely made up of a power-law distribution of pulses, the energy distribution should have a cut off around this value. However, this is unlikely as the observed GPs have a different ratio between the components C1 and C2 with respect to the integrated emission and have a different spectral index. We conclude that the power-law distribution of the energies of GPs must have a break or a cut off at lower energies as predicted by the emission model by \cite{Petrova2004} and observed in the Crab pulsar \citep{Karuppusamy2010}.

The time-of-arrival statistics of the GPs can be described with high accuracy with Poissonian statistics. This result, already discussed by \cite{Knight2007}, suggests that the GPs occur independently from each other on very small timescales. 

A total of 76 pulses show double GPs in the same rotation. In 74 cases both GPs occur in the component C1, while in 2 cases we observe GPs in component C1 and C2. This result is consistent with the assumption that each GP is independent form each other. A single example of a triple GP with two peaks in component C1 and one in C2 is also observed. The probability of registering one such case in the entire observation is only 1.4 percent which is still compatible at 2.5 $\sigma$ with the assumption above. An analysis of the phase differences between the double peaked GPs suggests that the phases of the GPs are not influenced by each other.

A number of theoretical model of GPs require sudden release of accumulated energy \citep{Istomin2004,Lyutikov2007,Machabeli2019}. This is at odds with the observational evidence, that suggests that the GPs occur independently from one another even within the same rotation and that there is no correlation between peak flux and wait times between the GPs. If there is an accumulation time, this must be shorter than the resolution of our observations of 9 $\mu$s, corresponding to less than 0.002 of the pulsar rotational phase.

%polarization
The GPs have low polarization percentages, comparable to the ones of the integrated emission. The average linear polarization of the integrated profile is only 1 percent, while the circular polarization is at 3 percent. Only 39 GPs show strong enough linear polarization to accurately determine the position angle (PA) which shows no correlation with phase. The low polarization levels are not in accordance with the induced scattering model of \cite{Petrova2004b,Petrova2004} which predicts high levels of polarization given that only the polarized ordinary waves of emission are focused and amplified in GPs. In 4 cases it was possible to measure the RM of the GPs which is highly variable and, in two cases, is not consistent with the RM of the total integrated pulsar signal. The levels of circular polarization in the GPs are higher, with 936 GPs showing circular polarization stronger than 5$\sigma$ the noise.
In double GPs, the polarization properties of the two peaks does not seem to be related, reinforcing the idea that they might be independent from one another. 

%observations in UHF band and with higher time resolution
The high rate and peculiar properties of the GPs in this pulsar make it also an interesting target for future studies. In particular, further observations of J1823$-$3021A with higher time resolution are necessary to resolve the internal structure of the GPs. That is not detected in our observation, but is a prediction of the model by \cite{Petrova2004}. 
%On the other hand, observations with higher sensitivity are needed to reconstruct the GP energy distribution at lower energies to look for the predicted deviations from a single power-law.
 Given the average spectral index of the GPs, a MeerKAT observation with the same total duration as ours, but using the UHF receivers and the entire array would be sensitive to GPs 3.5 times weaker than the weakest GPs reported in this paper, translating in a factor 8.5 increase in the detection rate. That would be essential to reconstruct the GP energy distribution at lower energies and then search for the predicted deviations from a single power-law. 

\section*{Acknowledgements}

The MeerKAT telescope is operated by the South African Radio Astronomy
Observatory, which is a facility of the National Research Foundation,
an agency of the Department of Science and Innovation. We thank Aris Karastergiou, Simon Johnston, and Robert Main for their comments and suggestions.
This research was funded partially by the Australian Government through the Australian Research Council, grants CE170100004 (OzGrav) and FL150100148. FA acknowledges the support, in the early stage of this work, from the Ministero degli Affari Esteri e della Cooperazione Internazionale - Direzione Generale per la Promozione del Sistema Paese - Progetto di Grande Rilevanza ZA18GR02. FA and MK gratefully acknowledge support from ERC Synergy Grant “BlackHoleCam” Grant Agreement Number 610058. AR gratefully acknowledges financial support by the research grant ``iPeska'' (P.I. Andrea Possenti) funded under the INAF national call Prin-SKA/CTA approved with the Presidential Decree 70/2016. 

\section*{Data availability}
The data will be made publicly available later this year on a portal under development on a OzSTAR supercomputer at Swinburne University of Technology . Early access to the raw data can be granted by application for an account on the OzSTAR supercomputer via https://supercomputing.swin.edu.au/account-management/.
%%%%%%%%%%%%%%%%%%%%%%%%%%%%%%%%%%%%%%%%%%%%%%%%%%

%%%%%%%%%%%%%%%%%%%% REFERENCES %%%%%%%%%%%%%%%%%%

% The best way to enter references is to use BibTeX:

\bibliographystyle{mnras}
\bibliography{GPs_biblio} % if your bibtex file is called example.bib

% Don't change these lines
\bsp	% typesetting comment
\label{lastpage}
\end{document}